\documentclass[twocolumn,showpacs,pre,aps,superscriptaddress,amsmath,amssymb,nofootinbib]{revtex4-1}

\usepackage[T1]{fontenc}
\usepackage{lmodern}
\usepackage{calrsfs}
\usepackage{upgreek}
\usepackage[mathscr]{eucal}

\usepackage{amsmath}

\usepackage{graphicx}

\usepackage{bm}
\usepackage{hyperref}

\graphicspath{{./Figures/}}

\begin{document}

\title{Maximum in density heterogeneities of active swimmers}

\author{Fabian Jan Schwarzendahl}
\affiliation{Max-Planck-Institute for Dynamics and Self-Organization, Am Fassberg 17, 37077 G\"ottingen, Germany}
\affiliation{Georg-August-Universit\"at G\"ottingen, Friedrich-Hund-Platz 1, 37077 G\"ottingen, Germany}
\author{Marco G. Mazza}
\affiliation{Max-Planck-Institute for Dynamics and Self-Organization, Am Fassberg 17, 37077 G\"ottingen, Germany}

\date{\today}

\begin{abstract}
Suspensions of unicellular microswimmers such as flagellated bacteria or motile algae  exhibit spontaneous density heterogeneities at large enough concentrations. 
Based on the relative location of the biological actuation appendages (\emph{i.e.} flagella or cilia) microswimmers' propulsion mechanism can be classified into two categories: (i) pushers, like \textit{E. coli} bacteria or spermatozoa, that generate thrust in their rear, push fluid away from them and propel themselves forward; (ii) pullers, like the microalgae \textit{Chlamydomonas reinhardtii}, that have  two flagella attached to their front, pull the fluid in and thereby generate thrust in their front. 
We introduce a novel model for biological microswimmers that creates the flow field of the corresponding microswimmers, and takes into account the shape anisotropy of the swimmer's body and stroke-averaged flagella. 
By employing multiparticle collision dynamics, we directly couple the swimmer's dynamics to the fluid's. 
We characterize the nonequilibrium phase diagram, as the filling fraction and P\'eclet number are varied, and find density heterogeneities in the distribution of both pullers and pushers, due to hydrodynamic instabilities. 
We find a maximum degree of clustering at intermediate filling fractions and at large P\'eclet numbers
resulting from a competition of hydrodynamic and steric interactions between the swimmers. We develop an analytical theory that supports these results.
This maximum might represent an optimum for the microorganisms' colonization of their environment.
\end{abstract}

\pacs{}

\maketitle

\section{Introduction}
\label{Sec:intro}

Physical interactions in suspensions of microswimmers consisting of bacteria or algae have been recognized to play an important role in the swimmers collective behavior \cite{BaskaranPNAS2009,EzhilanPoF2013,elgetiRepPRogPhys2015}. 
The nonequilibrium character of active suspensions, where the energy injection takes place at the scale of the microorganisms,
produces myriad mesmerizing phenomena, such as the spontaneous formation of spiral vortices~\cite{wiolandPRL2013},
directed motion~\cite{wiolandNJoP2016},
swarming~\cite{copelandSM2009}, bacterial turbulence~\cite{wensinkPNAS2012}, complex interaction with solid surfaces~\cite{laugaBiophysJ2006,berkePRL2008}, and self-concentration~\cite{dombrowskiPRL2004}.

Almost invariably, motile microorganisms move in an aqueous environment, where, because of their size,  viscous forces dominate, and inertial forces are completely negligible. In fact, consideration of the Navier--Stokes equations identifies that the nature of the dynamics is dictated by the ratio of viscous to inertial forces, known as the Reynolds number $\mathcal{R}=av\rho/\eta$, where $a$ is the typical size of the microorganism, 
$v$ its mean velocity, and $\rho$, $\eta$ are the fluid's density and viscosity, respectively. For \emph{Escherichia coli}, e.g., $a\approx 10~\upmu$m, $v\approx 30~\upmu$m/s, and for water $\rho\approx 10^3$ kg/m$^3$, $\eta\approx 10^{-3}$Pa s, which result in $\mathcal{R}\approx 10^{-5}$. 
As noted by Purcell~\cite{purcellAmJPhys1977}, this means that if the propulsion of a swimmer were to suddenly disappear, it would only coast for $0.1$ \AA. Thus, the state of motion  is only determined  by the forces acting at that very moment, and inertia is negligible.

 Due to the microswimmers' low Reynolds numbers,  the sum of viscous drag and thrust balances out to zero, in most situations. A direct consequence of force-free motion is that the leading term of the solution of the Stokes equation for a microswimmer is a symmetric force dipole (or stresslet).

%In the recent years the squirmer model \cite{GotzePRE2010} \cite{DowntonJPCM2009} 
%has shown great strength in modeling artificial microswimmers like colloids \cite{ZottelPRL2014} \cite{BlaschkeSM2016} \cite{AlarconSM2017} or droplets,
%as well as biological mircoswimmers like \textit{Volvox}. 

Biological microswimmers are complex systems because of the combination of biological, biochemical and physical processes all taking place at the same time. It is thus of great scientific value to develop theoretical models that isolate the relevant degrees of freedom and interactions. 
Considerable work has been done in recent years, and various models have been introduced, 
like the squirmer model \cite{lighthillComPApplMath1952,blakeJFM1971,IshikawaPRL2008,DowntonJPCM2009,GotzePRE2010,llopisJNNewtFlMech2010,evansPoF2011,ZottlPRL2012,alarconJMolLiq2013,IshimotoPRE2013,molinaSM2013,ZottlPRL2014,TheersSM2016,BlaschkeSM2016,AlarconSM2017,IshikawaPRE2010,LiPRE2014,PagonabarragaSM2013}, 
the shape anisotropic raspberry swimmer  \cite{deGraafJCP2016,fisherJCHEM2015,deGraafJCHEM2015},
the force-counterforce model \cite{nashPRL2010,nashPRE2008,hernandezPRL2005,StenhammmarPRL2017}, the catalytic dimers \cite{ValadaresSmall2010},
or other hydrodynamic models \cite{SaintillanPRL2007,SinghJSTM2015,SwanPF2011}.
Experiments have confirmed that the flow field of flagellated bacteria like \textit{E. coli} is to very good approximation modeled by a simple force dipole \cite{DrescherPNAS2011}, 
whereas \textit{Chlamydomonas} are modeled by three Stokeslets \cite{DrescherPRL2010}. 
Furthermore, as cell shapes vary greatly in the natural world, and realistic steric interactions are important in dense suspensions, a model that allows for flexibility in the shape of a microswimmmer is a highly desirable feature.
In this article we fill this lacuna. We derive a model for a flexible-shape microswimmer that produces self-propulsion by means of a force dipole for pusher-like microswimmers, or three Stokeslets for puller-like microswimmers. 

An efficient method to simulate fluids at mesoscopic scales, and their hydrodynamics is the multiparticle collision dynamics (MPCD) technique~\cite{MalevanetsJCP1999}. MPCD is a particle-based simulation method that correctly produces hydrodynamic modes. Due to its particle nature MPCD naturally includes thermal fluctuations, and can be easily  coupled to molecular dynamics methods of solutes, colloids~\cite{gompperBookChap2009}, and active swimmers~\cite{ZottlPRL2012,ZottlPRL2014,BlaschkeSM2016}. The MPCD technique in fact proves to be ideal for our purposes.

The nonequilibrium phase diagram of microswimmers has been subject to considerable interest, especially with regard to the emergence of density heterogeneities in the swimmers' distribution. 
We explore the phase diagram of active swimmers and 
show the presence of heterogeneities in the spatial distribution of both pushers and pullers. 
These heterogeneities arise due to the hydrodynamic interactions between the swimmers and relate to existing hydrodynamic theories
\cite{ishikawajfm2008,EzhilanPoF2013,SaintillanPRL2008,UnderhillPRL2008}. 
Interestingly, we find a maximum in the heterogeneities as filling fraction and P\'eclet number are varied. 
By using both computer simulations and analytical theory, we demonstrate
that this maximum results from a competition 
between hydrodynamic and steric interactions, where the latter suppress the hydrodynamic instability at higher filling fractions.
This optimum might have important biological implications on the ability of motile bacteria and algae to form colonies or biofilms. 

The remainder of this article is organized as follows. In Sec.~\ref{Sec:model} we introduce the model for the microswimmer,
the implementation of the fluid, and its coupling to the former. 
Section~\ref{Sec:flowfield} describes the physical properties of the fluid and the microswimmer's flow field. 
In Sec.~\ref{Sec:Denshetero} we present the nonequilibrium phase diagram of our model microswimmers, 
and specifically we characterize the density heterogeneities emerging from their hydrodynamic interactions
and show that these are suppressed by steric interactions at higher filling fractions.
In Sec.~\ref{Sec:Analytical} we present the analytical theory and show that we also 
find a maximum heterogeneity, which is mediated by the interplay of hydrodynamic and steric interactions. Finally, in Sec.~\ref{Sec:conclusion} we discuss our main results and summarize our conclusions.

\section{Model}
\label{Sec:model}

\begin{figure}[!htbp]
        \centering
        \includegraphics[width=0.9\columnwidth]{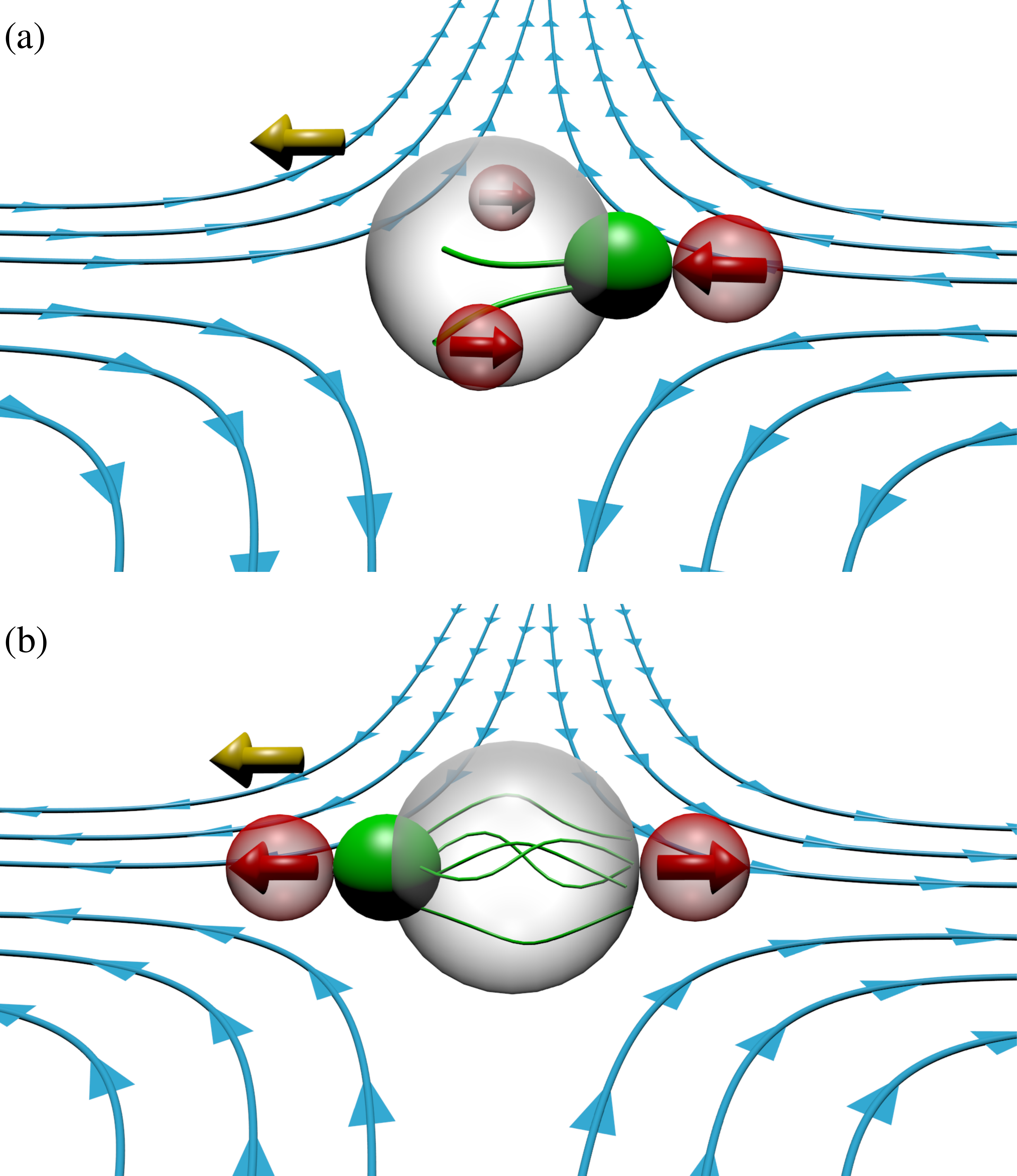}
        \caption{Schematic representation of the active swimmer model for (a) a puller type, and (b)  a pusher type microswimmer. 
        The small (green) sphere represents the swimmer's body, and the larger (transparent) sphere represents the stroke-averaged space spanned by the flagella. 
        The red spheres with embedded arrows represent the regions where the forces are applied. The golden arrows represent the swimming direction.
        The lines with arrows are a sketch of the streamlines.
       }
        \label{fig:SwimmerModel}
\end{figure}

We employ a stroke-averaged model of biological microswimmers, similarly to \cite{BaskaranPNAS2009,WysockiPRE2015}, taking into account the asymmetric shape of biological microswimmers due to the cell's body and the flagella. 
Thus, the swimmer is modeled as an asymmetric dumbbell, as depicted in Fig.~\ref{fig:SwimmerModel}, that
mimics a \textit{Chlamydomonas} or an \textit{E. coli} cell. The smaller sphere models the swimmer's body
and the larger sphere is a stroke average of the region spanned by flagellar motion.
We simulate the fluid surrounding the swimmer using MPCD (see Sec.~\ref{subsec:MPCalgorithm}).
Precise measurements~\cite{DrescherPNAS2011} show that pusher-type microswimmers are well modeled by a force dipole, represented by the red regions with embedded arrows in Fig.~\ref{fig:SwimmerModel}(b).
Since the swimmer is shape-asymmetric the hydrodynamic center is shifted with respect to the center 
of mass and henceforth a symmetry breaking on the swimmer gives propulsion. 
Pullers, on the other hand, are well represented by a three-Stokeslet solution of the Stokes equation \cite{DrescherPRL2010} 
[red regions in Fig.~\ref{fig:SwimmerModel}(a)].

The microswimmer is characterized by its mass $M$, center of mass
position $\bm{R}$ and orientation $\bm{q}$. The center of mass position in the swimmer's body frame and the inertia tensor $\mathbf{I}_\mathrm{m}$ are calculated in App.~\ref{subsec:CoM}. In the following we describe the equations governing the motion of the microswimmer, the fluid dynamics implemented through the MPCD, and their coupling.

\subsection{Rigid body dynamics of swimmers}

As we do not consider shape deformable swimmers, we are only concerned with rigid-body dynamics. 
The most general motion of a rigid body is the combination of a translation around an axis (the Mozzi axis) and a rotation around the same axis, as per the Mozzi--Chasles theorem~\cite{goldsteinBook}. 
Any orientation in space can be described using three numbers, that are commonly represented with the Euler angles, which correspond to three elementary rotations. However, there are a number of issues with the choice of the Euler angles. For instance, composition of rotations with Euler angles or rotation matrices is rather complex, and involves trigonometric functions which lead to an accumulation of rounding-off errors. Eventually the matrices representing the rotations may become not orthogonal. Importantly, for some values of the Euler angles there are discontinuous jumps in the representation. More fundamentally, the Euler angles do not generate a covering map of the rotation group SO(3), that is, the map from Euler angles to SO(3) is not always a local homeomorphism. Fortunately, the topology of SO(3) is diffeomorphic to the real projective space $\mathbb{P}_3(\mathbb{R})$ which admits a universal cover represented by the group of unit quaternions 
$\bm{q}= (q_0,q_1,q_2,q_3)^\mathrm{T}$,
where the superscript $\mathrm{T}$ indicates the matrix transposition. 
In App.~\ref{subsec:QuatFormulae} we provide the basic definitions of quaternion algebra necessary for the following.

The equations of motion for the rigid body dynamics in three dimensions read~\cite{OmelyanPRE1998}
\begin{align}
        m \ddot{\bm{R}} &= \bm{F} \,,
        \label{eq:Newton2}
        \\
        \ddot{\bm{q}} &= \frac{1}{2} \left[ \mathbf{W}\left( \dot{\bm{q}}\right) 
        \begin{pmatrix}
                0\\ \bm{\varOmega}^\mathrm{b}
        \end{pmatrix}
        +
          \mathbf{W}\left( \bm{q} \right) 
        \begin{pmatrix}
        0\\  
        \dot{\bm{\varOmega}}^\mathrm{b}
        \end{pmatrix}
        \right]\,, \\
        \dot{ \bm{q}} &= \frac{1}{2} \mathbf{W}(\bm{q}) 
        \begin{pmatrix}
                0\\ \bm{\varOmega}^\mathrm{b}
        \end{pmatrix}\,, \\
        \dot{\Omega}_{\alpha}^\mathrm{b} &= (I_\mathrm{m}^\mathrm{b})^{-1}_{\alpha} \left( T^\mathrm{b}_{\alpha} + \left( (I_\mathrm{m}^\mathrm{b})_{\beta} - (I_\mathrm{m}^\mathrm{b})_{\gamma} \right) \Omega_{\beta}^\mathrm{b} \Omega_{\gamma}^\mathrm{b} \right),
\label{eq:RigidBodyAngle}
\end{align}
where $\bm{\varOmega}$ is the angular velocity of the swimmer, $\mathbf{I}_\mathrm{m}^\mathrm{b}$ the moment of inertia tensor of the swimmer in the body frame, and
the indices $(\alpha,\beta,\gamma)$ take on as values the cyclic permutations of $(x,y,z)$. In Eq.~\eqref{eq:Newton2},
$\bm{F}=-\nabla \Phi$ and $\bm{T}= \bm{R}_\mathrm{F} \times \bm{F}$ are the force and 
torque, respectively, acting on the swimmer due to steric interactions with the neighbor, where $\bm{R}_\mathrm{F}$ is the vector connecting the center of mass of the 
swimmer to the point of contact with the neighbor, and the matrix $\mathbf{W}$ is given in App.~\ref{subsec:QuatFormulae}.
 The repulsive, steric interactions among swimmers are modeled using a Weeks--Chandler--Andersen potential \cite{WeeksJCP1971}
\begin{equation} \label{eq:WCAPotential}
  \Phi (r_{ij,ab}) = 
            4 \epsilon \left[ \left(\frac{\sigma_{ab}}{r_{ij,ab}} \right)^{12}-  
                              \left(\frac{\sigma_{ab}}{r_{ij,ab}} \right)^6 \right] 
            + \epsilon           
\end{equation}
if $r_{ij,ab} < 2^{1/6}\sigma_{ab}$, and $\Phi (r_{ij,ab}) = 0$ otherwise, 
where $r_{ij,ab}\equiv |\bm{r}_{ia}-\bm{r}_{jb}|$ is the distance between sphere $a$ of swimmer $i$ and sphere $b$ of swimmer $j$,
 $\epsilon$ is the energy scale and $\sigma_{ab}$ is the sum of the radii of sphere $a$ and sphere $b$.
For the numerical integration we use the Verlet algorithm proposed in \cite{OmelyanPRE1998}, which was also 
used and discussed in detail in \cite{TheersSM2016}.

Given a vector in the laboratory frame $\bm{f}$ the transformation to the body frame  vector $\bm{f}^\mathrm{b}$ is given by
\begin{align}
        \bm{f}^\mathrm{b} = \mathbf{D} \bm{f},
        \label{eq:BodyLabTrafo}
\end{align}
where the matrix $\mathbf{D}(\bm{q})$ is constructed from the quaternions and given in App.~\ref{subsec:QuatFormulae}.
Thus, the orientation of the swimmer at any given time is found from $\mathbf{D}^{-1}(\bm{q}(t))(0, 0, 1)^\mathrm{T}$

Note that all quantities that do not carry an index $b$ are calculated in the laboratory frame.

\subsection{Multiparticle collision dynamics}
\label{subsec:MPCalgorithm}

To simulate a fluid at fixed density $\rho$ and temperature $T$ surrounding the swimmers, we use the MPCD algorithm, which is a mesoscopic, particle based method \cite{MalevanetsJCP1999} to simulate a fluid at the Navier--Stokes level of description. We include the Anderson thermostat and the conservation of angular momentum into the MPCD dynamics; the resulting algorithm is usually denoted as MPC-AT+a~\cite{NoguchiEPL2007,GotzePRE2007,gompperBookChap2009}. 
The fluid is modeled using $N_\mathrm{fl}$ point-like particles of mass $m$, whose dynamics are executed through 
two steps: the streaming step and the collision step. In the streaming step the fluid particles positions $\bm{r}_i \,,\, i\in [1,N_\mathrm{fl}] $ are updated according to
\begin{equation}\label{eq:MPCaStream}
        \bm{r}_i (t + \delta t) =  \bm{r}_i (t) +  \bm{v}_i(t) \delta t,
\end{equation}
where $\bm{v}_i(t)$ is their velocity and $ \delta t $ is the MPCD timestep. 

The collision step mediates the interactions between the particles.
Here, the system is divided into $N_c$ collision cells with a regular grid of lattice constant $a$. 
The center of mass velocity in each cell $\mathsf{C}(i)$ is calculated and remains constant during the collision step, whereas the fluctuating part of the velocity of every fluid particle $i$ is randomized, which mimics the collision between particles. 
Hence, the velocity of  particle $i$ is updated as follows \cite{NoguchiEPL2007}
\begin{align}
        \bm{v}_i' = & \frac{1}{N_{\mathsf{C}(i)}} \sum_{j \in \mathsf{C}(i)} \bm{v}_{j}  +\bm{v}_i^{\text{ran}} - \frac{1}{N_{\mathsf{C}(i)}}\sum_{j \in \mathsf{C}(i)} \bm{v}_{j}^{\text{ran}}  \nonumber  \\
        &+ m  \left\lbrace\bm{\Pi}^{-1} \sum_{j \in \mathsf{C}(i)}\left[ \bm{r}_{j,c} \times \left( \bm{v}_i - \bm{v}_i^{\text{ran}}\right) \right] \right\rbrace\times \bm{r}_{i,c},
        \label{eq:MPCaCollision}
\end{align}
where the random velocity $\bm{v}_i^{\text{ran}}$ has components distributed according to a Gaussian distribution with zero mean and variance $\sqrt{k_\mathrm{B}T/m}$, $k_\mathrm{B}$ is the Boltzmann constant, 
 $N_{\mathsf{C}(i)}$ is the number of fluid and ghost particles (see Sec.~\ref{subsec:coupling}) in cell $\mathsf{C}(i)$.
The vector $\bm{r}_{j,c}$ is the position of the neighboring particle $j$ relative
to the center of mass of the cell $\mathsf{C}(i)$. In Eq.~\eqref{eq:MPCaCollision},
$\bm{\Pi}^{-1}$ is the inverse of the moment of inertia tensor $ \bm{\Pi}\equiv \sum_{j \in \mathsf{C}(i)}m\left[(\bm{r}_j\cdot\bm{r}_j){\mathbf{I}}-\bm{r}_j\otimes\bm{r}_j\right]$ for the fluid particles in cell $\mathsf{C}(i)$, where $\mathbf{I}$ is the identity tensor, `$\cdot$' is the scalar product and `$\otimes$' the tensor product. 
Note that $\bm{\Pi}^{-1}$ is a dynamical quantity that has to be updated at every timestep, and
it also includes the ghost particles within the swimmer (see Sec.~\ref{subsec:coupling}).

To ensure Galilean invariance and avoid the build-up of spurious correlations in the velocities~\cite{IhlePRE2001}, the usual grid shift is performed at each timestep, that is, the grid is shifted by a random vector, whose components are uniformly distributed in the interval $[-a/2, a/2]$.

\subsection{Coupling of the swimmer's and fluid's dynamics}\label{subsec:coupling}

No velocity field is prescribed in our model of microswimmers. Locomotion is achieved by obeying the conservation of momentum in the collisions between the fluid particles and the swimmers; the shape asymmetry then induces self-propulsion.
Two physical effects need to be included: we impose no-slip boundary conditions on the model swimmer's surface, and the force poles are explicitly included (see Fig.~\ref{fig:SwimmerModel}). 
Both effects induce modifications of the streaming and collision steps of the MPCD algorithm that we explain in the following.

\subsubsection{Streaming step}
To ensure the no-slip boundary condition the bounce-back rule \cite{LamuraEPL2001} 
is applied to the MPCD particles that hit the spheres during the streaming step.
The velocity of the fluid particle is reversed and the change in momentum is given by
\begin{align}
        \bm{J}_i = 2 m \left( \bm{v}_i - \bm{U} - \bm{\varOmega} \times \left( \bm{{\tilde{r}}}_i - \bm{R} \right)\right),
        \label{eq:BounceBackLinMomentumChange}
\end{align}
where $\bm{R}$  is the center-of-mass position of the swimmer colliding with the fluid particle,
$\bm{U}$ and $\bm{\varOmega}$ are the linear and angular velocity of the swimmer, $\bm{{\tilde{r}}}_i$ is the position of the fluid particle upon collision with the sphere. 
The updated fluid velocity reads
\begin{align}
        \bm{v}'_i= \bm{v}_i - \bm{J}_i/m.
        \label{eq:MPCpartBounceBack}
\end{align}
In addition, the fluid particles are reflected back along the direction of their initial velocity.
For this, we use an exact ray tracing method to detect the collision of the MPCD particle onto the 
swimmer's surface.
If a collision is detected the MPCD particle is propagated back onto the swimmers surface and
then the bounce-back rule is applied. 
The new linear and angular velocities of the swimmer after the collision with the fluid particles read
\begin{align}
        \bm{U}' &= \bm{U} + \sum_i \bm{J}_i /M
        \\
        \bm{\varOmega}' &= \bm{\varOmega} + \mathbf{I}_m^{-1} \sum_i (\bm{r}_i - \bm{R}) \times \bm{J}_i\,.
        \label{eq:SwimmerBounceBack}
\end{align}

The force poles are added as external force regions (see \cite{BolintineanuPRE2012} for external force) in the streaming step for each swimmer.
This is done by modifying the streaming step inside the force regions to
\begin{align}
        \bm{r}_i (t + \delta t) =  &\bm{r}_i (t) +  \bm{v}_i(t) \delta t + \bm{f}_\mathrm{lab}\frac{\delta t^2}{2}  ,  
        \label{eq:ModifiedStreamingPosition}\\
        \bm{v}_i (t + \delta t) =  &\bm{v}_i (t) +  \bm{f}_\mathrm{lab} \delta t,
        \label{eq:ModifiedStreamingVelocity}
\end{align}
where the force in the lab frame reads
\begin{align}
        \bm{f}^\mathrm{lab}_{\mathrm{ac}}\equiv \bm{f}_{\mathrm{ac}} - \left(  \bm{U}  + \bm{\varOmega} \times \left( \bm{r}_i - \bm{R} \right)\right) / \delta t,
        \label{eq:activeForce}
\end{align}
and $\bm{f}_{\mathrm{ac}}$ is the active force discussed in the following.
The flow fields are modeled by a force poles. While mathematically such force poles are point forces, any numerical implementation must mollify this requirement.
\begin{figure}[]
        \centering
        \includegraphics[width=1\columnwidth]{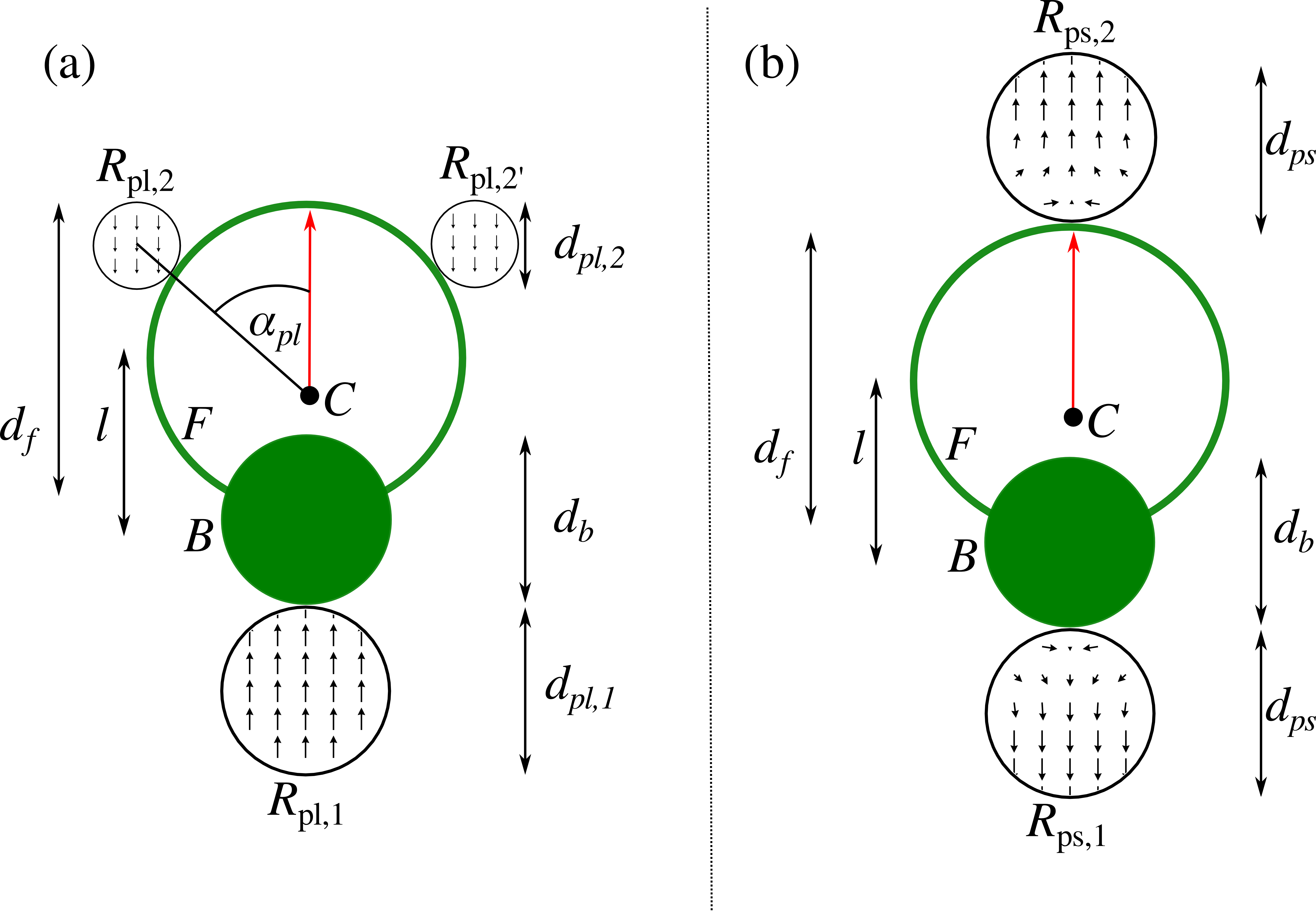}
        \caption{Geometry of the puller type swimmer (a) and pusher type swimmer (b). Regions $B$ (green) and $F$ (empty circle) are the body with diameter $d_b$, and stroke averaged flagella with diameter $d_f$, respectively,
                and they are separated by a distance $l$.
                The red arrow denotes the swimmers orientation and $C$ is the center of mass. 
                Black circles are the force poles acting on the fluid. For pullers (a) the region $R_{\mathrm{pl,1}}$ has the diameter $d_{\mathrm{pl,1}}$ and the regions $R_{\mathrm{pl,2}}$, $R_{\mathrm{pl,2'}}$ have the diameter $d_{\mathrm{pl,2}}$. 
                For pushers (b) the regions $R_{\mathrm{ps,1}}$ and $R_{\mathrm{ps,2}}$ have the diameter $d_{\mathrm{ps}}$. }
        \label{fig:swimmerGeometry}
\end{figure}

\emph{Pullers} -- The flow field is modeled by three Stokeslets, and the active force $\bm{f}_{\mathrm{ac}}^{\mathrm{lab}}$ is applied to $R_{\mathrm{pl,1}}$, $R_{\mathrm{pl,2}}$ and $R_{\mathrm{pl,2'}}$ [see Fig.~\ref{fig:swimmerGeometry}(a)]. 
The region $R_{\mathrm{pl,1}}$  with diameter $d_{\text{pl,1}}$ is located at the rear of the swimmer and its force points into the 
direction of the swimmers orientation. The other two regions $R_{\mathrm{pl,2}}$ and $R_{\mathrm{pl,2'}}$ are aligned on the side of the swimmer and have the opposite orientation. 
The angle $\alpha_{\text{pl}}$ between the orientation of the puller and the line connecting the center of mass $C$ and the midpoint of the region $R_{\mathrm{pl,2}}$ (or $R_{\mathrm{pl,2'}}$)  
defines their position on the boundary of the swimmer.
The diameter of both $R_{\mathrm{pl,2}}$ and $R_{\mathrm{pl,2'}}$ is $d_{\text{pl,2}} = d_{\text{pl,1}}/(2)^{1/3}$, such that they have half the volume of $R_{\mathrm{pl,1}}$, making the fluid force free.

\emph{Pushers} -- The flow field is modeled by a force dipole. We apply the force $\bm{f}_{\mathrm{ac}}^{\mathrm{lab}}$
to all fluid particles located within spherical regions [see Fig.~\ref{fig:swimmerGeometry}(b)]. %(  and in the lab frame). 
The regions $R_{\mathrm{ps,1}}$ and $R_{\mathrm{ps,2}}$ where $\bm{f}^\mathrm{lab}_{\mathrm{ac}}$ is applied are equally sized spheres with diameter $d_\mathrm{ps}$ and 
the two forces $\bm{f}_{\mathrm{ac}}$ 
are equal and opposite, to ensure that the fluid is overall force free. To generate a smooth flow on the boundary of the swimmer the direction of the applied force 
in regions $R_{\mathrm{ps,1}}$ and $R_{\mathrm{ps,2}}$ is modeled as follows.
For fluid particles $\bm{r}_i \in R_{\mathrm{ps,1}}$ or $R_{\mathrm{ps,2}}$ we apply the force
\begin{align}
        \bm{f}^b_{ac}= 
          \begin{cases}
                  \begin{pmatrix}0\\0\\1\end{pmatrix} f_0 , & \text{if } s_z^b > d_\mathrm{ps}/2,\\
                  \begin{pmatrix} \frac{2 s_z^b}{d_{\mathrm{ps}}} s_x^b/|\bm{s}^b| \\ \frac{2 s_z^b}{d_{\mathrm{ps}}} s_y^b/|\bm{s}^b|\\ (\frac{2 s_z^b}{d_{\mathrm{ps}}} -1) s_z^b/|\bm{s}^b|\end{pmatrix} f_0 , & \text{if } s_z^b < d_\mathrm{ps}/2.
           \end{cases}
        \label{eq:ModifiedForce}
\end{align}
Here, $\bm{s}^b = (s_x^b, s_y^b, s_z^b)^\mathrm{T}$ is the distance between the MPCD particle and the center of the region $\bm{r}_i \in R_{\mathrm{ps,1}}$ or $R_{\mathrm{ps,2}}$.
As before the superscript $b$ denotes the body frame, in which 
the swimmers orientation is aligned with the $z$ axis. 
The constant $f_0$ gives the strength of the force that is applied and $f_0 <0 $ in $R_{\mathrm{ps,1}}$ and $f_0 >0 $ in $R_{\mathrm{ps,2}}$.

\subsubsection{Collision step}

To guarantee the no-slip boundary conditions on the surface of the swimmers, it is necessary to fill 
each swimmer with ghost particles, such that the collision step can be properly executed \cite{GotzePRE2007}. 
The positions of the ghost particles $\bm{r}_i^g$ are uniformly 
distributed within the swimmer\footnote{We recommend to fill the swimmer with multiple ghost particles, rather than using a single ghost particle of large mass, as this would result in a wrong value of the torque.},
and are advected with the swimmer in each timestep. 
The ghost particles density is matched to the fluids density so as to make the swimmer neutrally buoyant.
%Note that it is computationally simplest to keep a list of which particles are ghosts and to keep a second position array, that saves the relative positions to the according swimmer. 
Before every collision step the ghost velocities $\bm{v}_i^g$ are updated according to
\begin{align}
        \bm{v}_i^g = \bm{U}  + \bm{\varOmega} \times \left( \bm{r}_i^g - \bm{R} \right) + \bm{v}_i^{\text{ran}},
        \label{eq:GhostInitialv}
\end{align}
where the components of $\bm{v}_i^{\text{ran}}$ are sampled from a Gaussian distribution. 
The ghost particles then (together with the fluid particles) take part in the collision step [see Eq.~\eqref{eq:MPCaCollision}], and their velocities are updated to $\bm{v}_i^{g\prime}$.
The resulting change in linear momentum due to the ghost particles is $\bm{J}_i^g = m\left( \bm{v}_i^{g\prime} -\bm{v}_i^g \right)$ and the change in angular momentum is $\bm{L}_i^g = \left( \bm{r}_i^g - \bm{R}  \right) \times \bm{J}_i^g$.
These changes are then transferred to the swimmer \cite{GotzePRE2010} 
\begin{align}
         \bm{U}' &= \bm{U} + \sum_i \bm{J}_i^g /M\,,
        \\
        \bm{\varOmega}' &= \bm{\varOmega} + \mathbf{I}_\mathrm{m}^{-1} \sum_i \bm{L}_i^g\,.
        \label{eq:SwimmerGhostMomentumChange}
\end{align}

To conclude this section on our computational model, we note that this algorithm scales as $\mathscr{O}(N)$, and thus is particularly prone to an efficient implementation with parallel programming. We therefore implemented the entire dynamics on graphics processing unit (GPU) cards.

\subsection{Computational details}

We carried out three-dimensional simulations with an average of $\langle N_\mathsf{C} \rangle = 20$ fluid particles per cell.
%for pusher type swimmers, and
%$\langle N_\mathsf{C} \rangle = 5$ for puller type swimmers.
The timestep of the MPCD algorithm is fixed to $\delta t = 10^{-2} \sqrt{m a^2/(k_{\mathrm{B}} T)}$, 
whereas the MD timestep is $\delta t_{\mathrm{MD}} = 5 \times 10^{-4} \sqrt{m a^2/(k_{\mathrm{B}} T)}$. 
%Note that a very small timestep is necessary to ensure that the fluid is incompressible.
The resulting kinematic viscosity $\nu=\eta/\rho$ of the fluid for the MPC-AT+a algorithm (including both kinetic and collisional contribution) can be calculated exactly as 
$\nu = 3.88 a \sqrt{k_\mathrm{B} T/m}$~\cite{GotzePRE2007,noguchiPRE2008,gompperBookChap2009}. 
Simulations using a forced flow (for details see \cite{BolintineanuPRE2012}) produced a viscosity of $\nu = 3.69  a \sqrt{k_\mathrm{B} T/m}$.
%for $\langle N_\mathsf{C} \rangle = 20$ and $\nu = 3.1 a \sqrt{k_\mathrm{B} T/m}$ for $\langle N_\mathsf{C} \rangle = 5$.

The large sphere $F$ associated to the stroke-averaged flagella of the swimmer has a diameter of $d_{\text{f}}= 7 a$, while the small sphere $B$ associated to the body of the swimmer has  $d_{\text{b}}= 3 a$, and 
the distance between the spheres centers is $l = 7 a$. 
The choice of the geometrical parameters is dictated by a combination of factors. First, it is computationally convenient to make the swimmers' linear size a few times the grid spacing $a$. 
Second, inspired by the geometric properties of \emph{Chlamydomonas} a ratio $d_{\text{f}}/d_{\text{b}}\gtrsim 2$ is advisable \cite{OstapenkoarXiv2016}. For the sake of clarity in the comparison of our results, we maintain the same geometry also for pushers. 
The energy scale of the steric interactions is set to $\epsilon= 10 k_{\mathrm{B}} T$.
For pushers we fix the diameter of the force dipole regions $R_{\mathrm{ps,1}}$ and $R_{\mathrm{ps,2}}$ to $d_{\text{ps}} = 3 a$. 
The region $R_{\mathrm{pl,1}}$ of the pullers has the same diameter $d_{\text{pl,1}} = 3 a$ and accordingly the regions $R_{\mathrm{pl,2}}$ and $R_{\mathrm{pl,2'}}$ have the diameter $d_{\text{pl,2}} = 3 a/(2)^{1/3}$.
The angle between the swimmers orientation and the line connecting the center of mass of the pullers $C$ to the midpoint of the regions $R_{\mathrm{pl,2}}$ and $R_{\mathrm{pl,2'}}$ is $\alpha_{\text{pl}}=  0.31$.

To initialize the simulations, we distribute the swimmers homogeneously across a cubic box with periodic 
boundary conditions.

\section{Characterization of the fluid and active hydrodynamics}\label{Sec:flowfield}

In the following we describe calculations aimed at characterizing the thermal (equilibrium) properties of our model in the passive case, and also the flow field generated by the active motion.

We first consider a passive colloid (with the same geometry described above) immersed in the MPCD fluid, that is, 
we carried out equilibrium simulations without activity $\bm{f}_{\text{ac}}^{\mathrm{lab}}=0$. 
The equipartition theorem applied to the passive colloid for the translational and rotational motion predicts
$\langle U_{\alpha}^2 \rangle  = {k_\mathrm{B} T}/{M}$, 
$\langle (\varOmega^\mathrm{b}_{\alpha})^2 \rangle  = {k_\mathrm{B} T}/{I_{\mathrm{m}\,\alpha}}$. 
For our system, we find a theoretical value of the translational motion $\langle U_{\text{theory}}^2 \rangle  = 2.7 \times 10^{-4} {k_\mathrm{B} T}/{m}$ 
and the simulations give 
$\langle U_{x}^2 \rangle  = \langle U_{y}^2 \rangle  =\langle U_{z}^2 \rangle  = 2.3 \times 10^{-4} {k_\mathrm{B} T}/{m}$.
The prediction for the angular motion in $x$ and $y$ direction yields $\langle (\varOmega^\mathrm{b}_{x,y})^2 \rangle  = 1.3 \times 10^{-3} {k_\mathrm{B} T}/{m a^2}$
while the simulations give $\langle (\varOmega^\mathrm{b}_{x})^2 \rangle  =\langle (\varOmega^\mathrm{b}_{y})^2 \rangle  = 1.1 \times 10^{-3} {k_\mathrm{B} T}/{m a^2}$.
In the $z$ direction, the theory predicts $\langle (\varOmega^\mathrm{b}_{z})^2 \rangle  = 1.4 \times 10^{-3} {k_\mathrm{B} T}/{m a^2}$
and the simulations give $\langle (\varOmega^\mathrm{b}_{x})^2 \rangle  = 1.2 \times 10^{-3} {k_\mathrm{B} T}/{m a^2}$.

\begin{figure*}[!htbp]
        \centering
        \includegraphics[width=2\columnwidth]{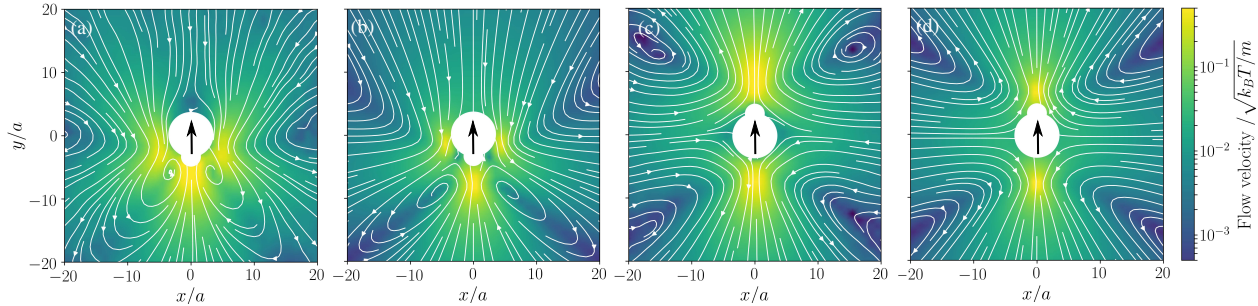}
        \caption{Time-averaged flow field generated by (a) our model puller, (b) theoretical puller, (c) our model pusher, and (d) theoretical pusher.
                We show cross-sections on the $x$-$y$ plane at $z=0$. The force strength is $f_0 =50 k_{\mathrm{B}} T/a$. The large central white regions show the active swimmers. 
        The thin lines with arrows mark the streamlines, while the color code shows the magnitude of the flow velocity normalized to the thermal velocity. The large black arrows indicate the direction of motion.}
        \label{fig:FlowSwimmer}
\end{figure*}

Hydrodynamics at low Reynolds numbers (relevant for micron-sized objects) allows a great simplification of the Navier--Stokes equations: the nonlinear, inertial effects can be neglected, and the governing equations are the Stokes equations
\begin{align}\label{eq:Stokes}
\eta\nabla^2\bm{v}=\nabla p-\bm{f}^{\mathrm{ext}}\,, \quad\quad  \nabla\cdot\bm{v}=0\,,
\end{align}
where $\bm{v}(\bm{r})$ is the fluid velocity,  $p(\bm{r})$ the pressure, and $\bm{f}^{\mathrm{ext}}(\bm{r})$ is a body force acting on the fluid. 
Solving the Stokes equations means obtaining expressions for $\bm{v}$ and $p$ that satisfy Eq.~\eqref{eq:Stokes} and the boundary conditions. From this knowledge, the stress tensor $\bm{\sigma}$ can be calculated. For a Newtonian fluid, $\bm{\sigma}$ depends linearly on the instantaneous values of the velocity gradient, so that one can write 
$\bm{\sigma}=-p\bm{\mathrm{I}}+\eta\left[\nabla \otimes \bm{v}+(\nabla \otimes \bm{v})^{\,\mathrm{T}}\right]$.
Because the Stokes equations are linear, their solution can be formally written in terms of the convolution of a Green's function with the inhomogeneous term $\bm{f}^{\mathrm{ext}}$~\cite{spagnolieJFM2012,zottlJPCM2016}
\begin{equation}
\bm{v}(\bm{r},t)=\int \mathbf{O}(\bm{r}-\bm{r'})\bm{f}^{\mathrm{ext}}(\bm{r'},t)\,d\bm{r'}\,. 
\end{equation}
In free three-dimensional space, the Greens function is found by considering a point force $\bm{f}^{\mathrm{ext}}=f\bm{e}\,\delta(\bm{r})$ in an infinite, quiescent fluid, where $\bm{e}$ is the unit vector representing the direction of the force. A straightforward calculation~\cite{dhont1996book} gives the Oseen tensor
$\mathbf{O}(\bm{r})\equiv \frac{1}{8\pi\eta r}\left( {\mathbf{I}}+{\bm{\hat{r}}\otimes\bm{\hat{r}}} \right)$  where $\bm{\hat{r}}\equiv \bm{r}/r$, $r=|\bm{r}|$, 
and the resulting flow field $\bm{v}(\bm{r})=\frac{f}{8\pi\eta r} \left[ \bm{e}+ (\bm{{\hat{r}}}\cdot\bm{e})\bm{\hat{r}} \right]$, which is termed a `Stokeslet' and decays with distance as $r^{-1}$.

A theoretical prediction for the puller flow field is constructed from three Stokeslets, and for the pusher we use two Stokeslets. 
The Stokeslets positions are placed at the midpoints of the respective force regions from the simulations.
For pushers, the force in one of the two regions can be estimated by integrating Eq.\eqref{eq:ModifiedForce}, which yields $f= f_0 \frac{5}{48}\pi d_{\text{ps}}^3 \rho$. 
This takes into account the redirection of 
the force on the boundary of the swimmer and the density $\rho$ of the fluid. 
For pullers the force in the region $R_{\mathrm{pl,1}}$ is $f = f_0 \frac{1}{6} \pi d_{\text{pl,1}}^3 \rho$ and in the regions $R_{\mathrm{pl,2}}$, $R_{\mathrm{pl,2'}}$ is $f = f_0 \frac{1}{6} \pi d_{\text{pl,2}}^3 \rho$.

We now consider the flow field generated by the active motion of our model microswimmer. We switch on the 
active motion with a force $f_0 = 50 k_{\mathrm{B}} T/a$
and carry out the full dynamics as described in Sec.~\ref{Sec:model}. Figure~\ref{fig:FlowSwimmer} shows the flow fields of a pusher and a puller in the  lab frame.
As expected, the flow field of the puller is contractile, as fluid is drawn in from the front and the back, while fluid is pushed away normal to the swimming direction [Fig.~\ref{fig:FlowSwimmer}(a)]. 
The situation is reversed in the pusher case [Fig.~\ref{fig:FlowSwimmer}(c)] where fluid is pushed out at the front and back of the swimmer.
The theoretical predictions for both the puller [Fig.~\ref{fig:FlowSwimmer}(b)] and the pusher [Fig.~\ref{fig:FlowSwimmer}(d)] show the same characteristic behavior as the simulated flow fields.

\begin{figure}[!htbp]
\includegraphics[width=1\columnwidth]{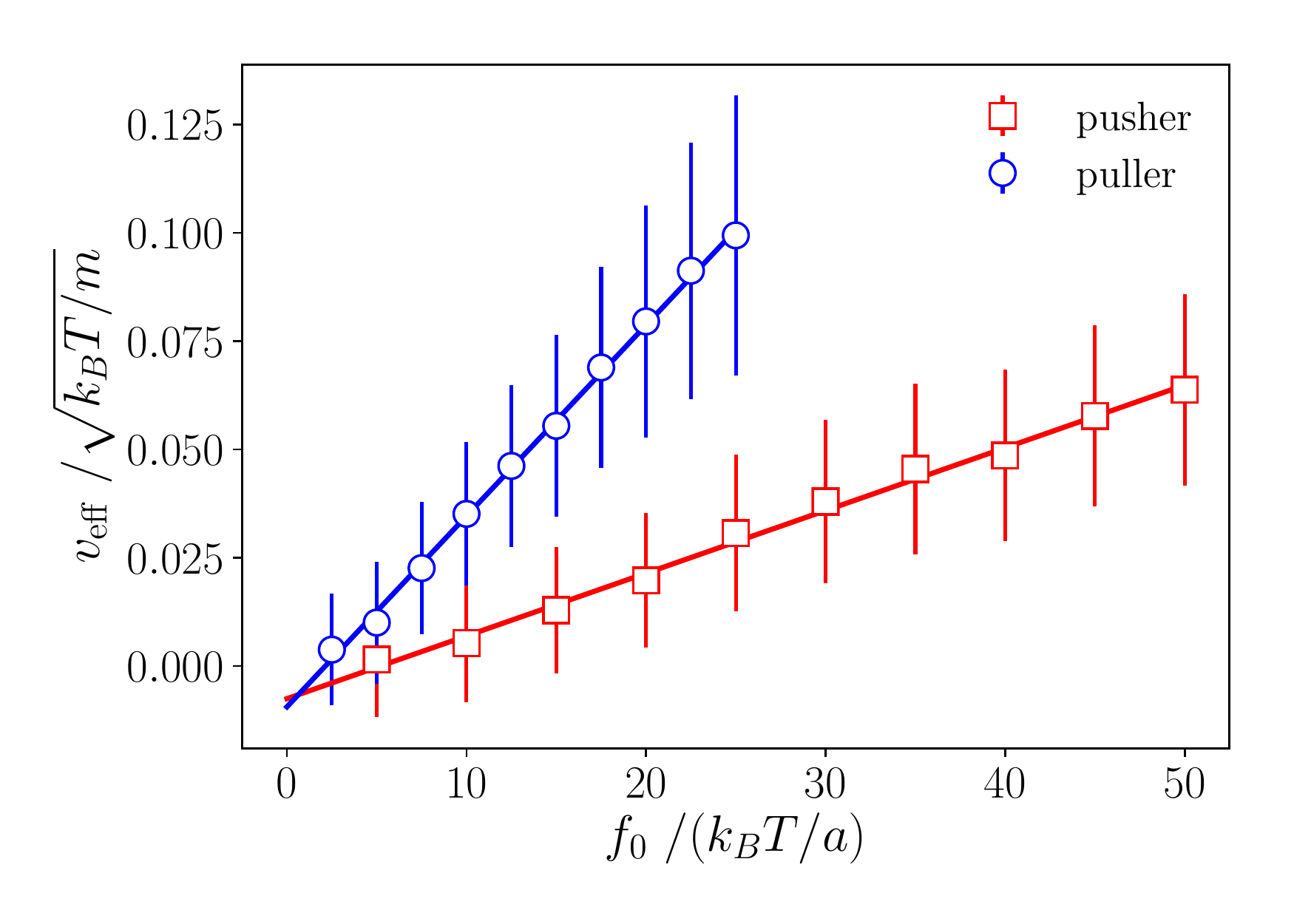}
\caption{Dependence of the pusher and puller velocity $v_{\mathrm{eff}}$ on the active force $ f_0 $.
                Lines are linear fits to the simulated data.}
        \label{fig:VelSwim}
\end{figure}

The effective velocity $v_{\mathrm{eff}}\equiv |\langle \bm{e} \cdot \bm{U} \rangle |$ of an isolated swimmer in the steady state depends linearly on the active force $f_0$~\cite{BaskaranPNAS2009}. From our simulations, we calculate $v_{\mathrm{eff}}$
for a pusher [see Fig.~\ref{fig:VelSwim}]. The linear fit has a slope of $\alpha= ( 1.45  \pm 5 \times 10^{-2}) \times 10^{-3}\sqrt{\frac{m a^2}{k_{\mathrm{B}} T}} $. 
The analogous results for pullers are also shown in Fig.~\ref{fig:VelSwim}, where the slope of the linear fit is $\alpha= ( 4.4  \pm 7 \times 10^{-2}) \times 10^{-3} \sqrt{\frac{m a^2}{k_{\mathrm{B}} T}} $.
For both pushers and pullers the maximal effective velocity we investigate is $v_{\mathrm{eff}} \sim 0.1 \sqrt{k_{\mathrm{B}} T/m}$, 
therefore the highest Reynolds number is $Re=\frac{ f_0 \alpha \sigma}{\nu} \sim 0.1$, and thus we are in the low Reynolds number regime.

\section{Density heterogeneities}
\label{Sec:Denshetero}

We carry out simulations with $N=300 -1560$ swimmers and study the phase 
behavior of the active swimmers. 
The filling fraction we denote here is computed using the volume of the swimmers body $V_{\text{B}}$,
as well as the volume that is spanned by the flagella sphere $V_{\text{F} }$, while taking into account their overlap volume $V_{\text{Ol}}$
\begin{align}
        \phi = \frac{(V_{\text{B}} + V_{\text{F}} -V_{\text{Ol}}  ) N}{V_{\text{Box}}}.
        \label{eq:fillingfrac}
\end{align}
Here $V_{\text{Box}}$ is the volume of the simulation box.
Note, that in an experiment only the body of the cell would be taken into account, 
thus the filling fraction should then be rescaled by $V_{\text{B}}/(V_{\text{B}} + V_{\text{F} }  -V_{\text{Ol}}) = 7.5 \times 10^{-2}$.
Furthermore, we vary the strength of the active force $ f_0 $, which changes the propulsion speed $v_{\mathrm{eff}}$ 
as well as the strength of the hydrodynamic interactions between the swimmers.
The P\'eclet number captures the ratio of advection to diffusion, and can be computed using
\begin{align}
        \mathcal{P}=  \frac{ f_0 \alpha \sigma}{D},
        \label{eq:PecletNumber}
\end{align}
where we used the linear relation (fitted slope) between active velocity and force dipole strength from Sec.~\ref{Sec:flowfield}.
Furthermore, $\sigma = 5a$ is the typical length of the swimmer and for the diffusion constant we 
assume $D = \frac{k_{\mathrm{B}} T}{ 6 \pi \eta \sigma}$.

\begin{figure}[!htbp]
\includegraphics[width=1\columnwidth]{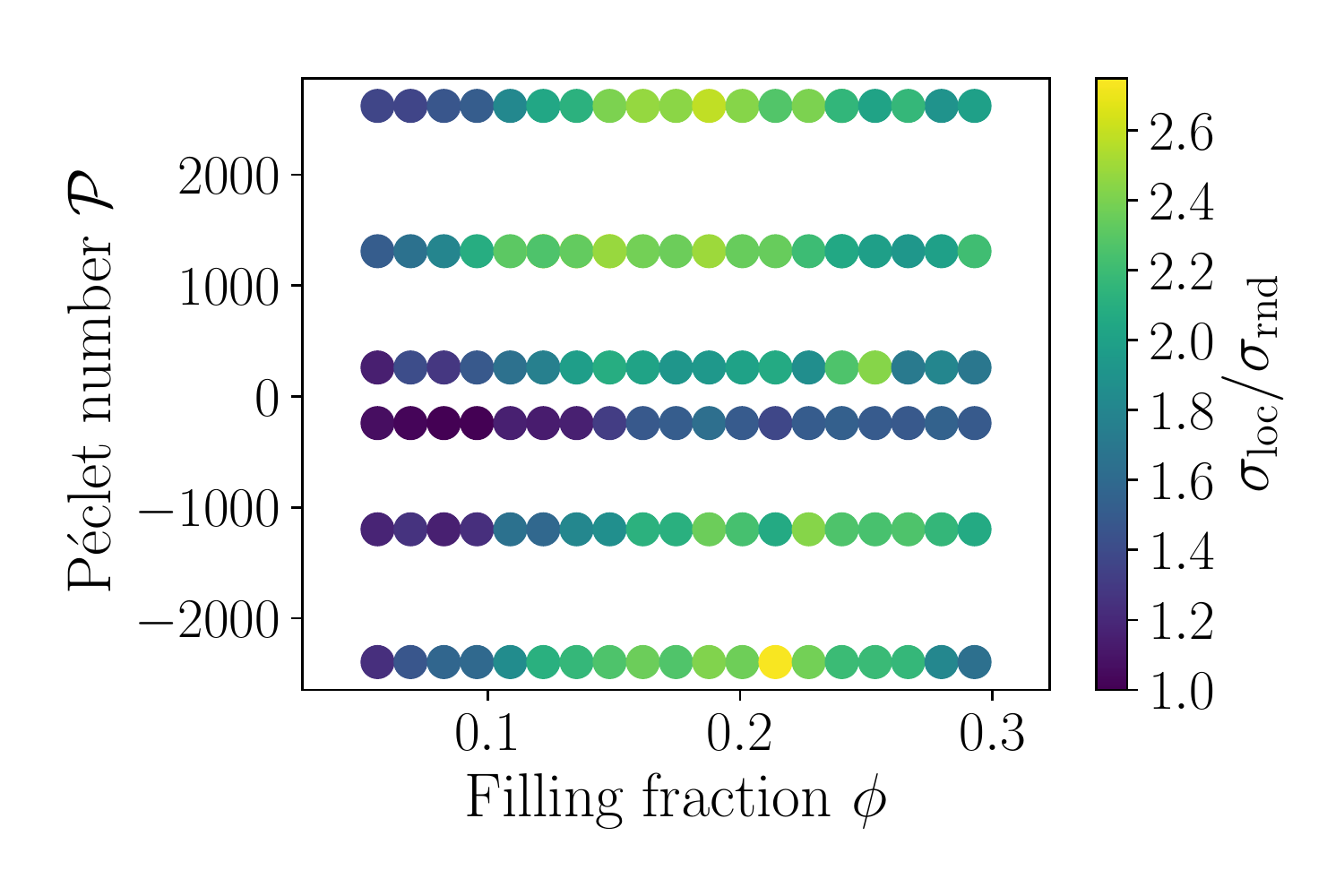}
        \caption{Standard deviation of local Voronoi volume $\sigma_{\mathrm{loc}}$ compared to 
                standard deviation $\sigma_{\mathrm{rnd}}$ of a homogeneous configuration. 
                The P\'eclet number $\mathcal{P}$ as well as the global filling fraction $\phi$ are varied.
                Positive P\'eclet numbers correspond to pusher type and negative to puller type swimmers.}
        \label{fig:InhomoDensPhasemap}
\end{figure}

We analyze the systems density using a Voronoi tessellation and compute the local volume for each swimmer. 
A global measure for the heterogeneity of a configuration of swimmers is given by the standard deviation of the
distribution of local Voronoi volumes $\sigma_{\mathrm{loc}}$. 
We compare $\sigma_{\mathrm{loc}}$ to the standard deviation of local Voronoi volumes for random homogeneous configurations $\sigma_{\mathrm{rnd}}$ of the 
corresponding filling fraction.
Figure~\ref{fig:InhomoDensPhasemap} shows the resulting phase diagram, where the ratio $\sigma_{\mathrm{loc}}/\sigma_{\mathrm{rnd}}$ as a function of P\'eclet number 
and filling fraction is shown.
Here, positive values of the P\'eclet number correspond to pusher type swimmers, whereas negative values are puller swimmers. 

The phase diagram shows that for both pullers and pushers,
initially $\sigma_{\mathrm{loc}}/\sigma_{\mathrm{rnd}}$ grows with P\'eclet number and filling fraction, then it reaches a maximum and drops to lower values. 
The initial increase is related to an instability that is mediated by the hydrodynamic interactions of the
microswimmers which has also been found in \cite{BaskaranPNAS2009, ishikawajfm2008, 
EzhilanPoF2013, SaintillanPRL2008, UnderhillPRL2008}.
As the filling fraction increases, steric interactions grow in importance and compete with the hydrodynamic instability.
Thus, we ascribe the presence of the maximum in $\sigma_{\mathrm{loc}}/\sigma_{\mathrm{rnd}}$ to a tapering effect of the steric interactions
on the hydrodynamic instability. This tapering effect becomes visible only when steric interactions are fully accounted for.
\begin{figure}[!htbp]
\includegraphics[width=1\columnwidth]{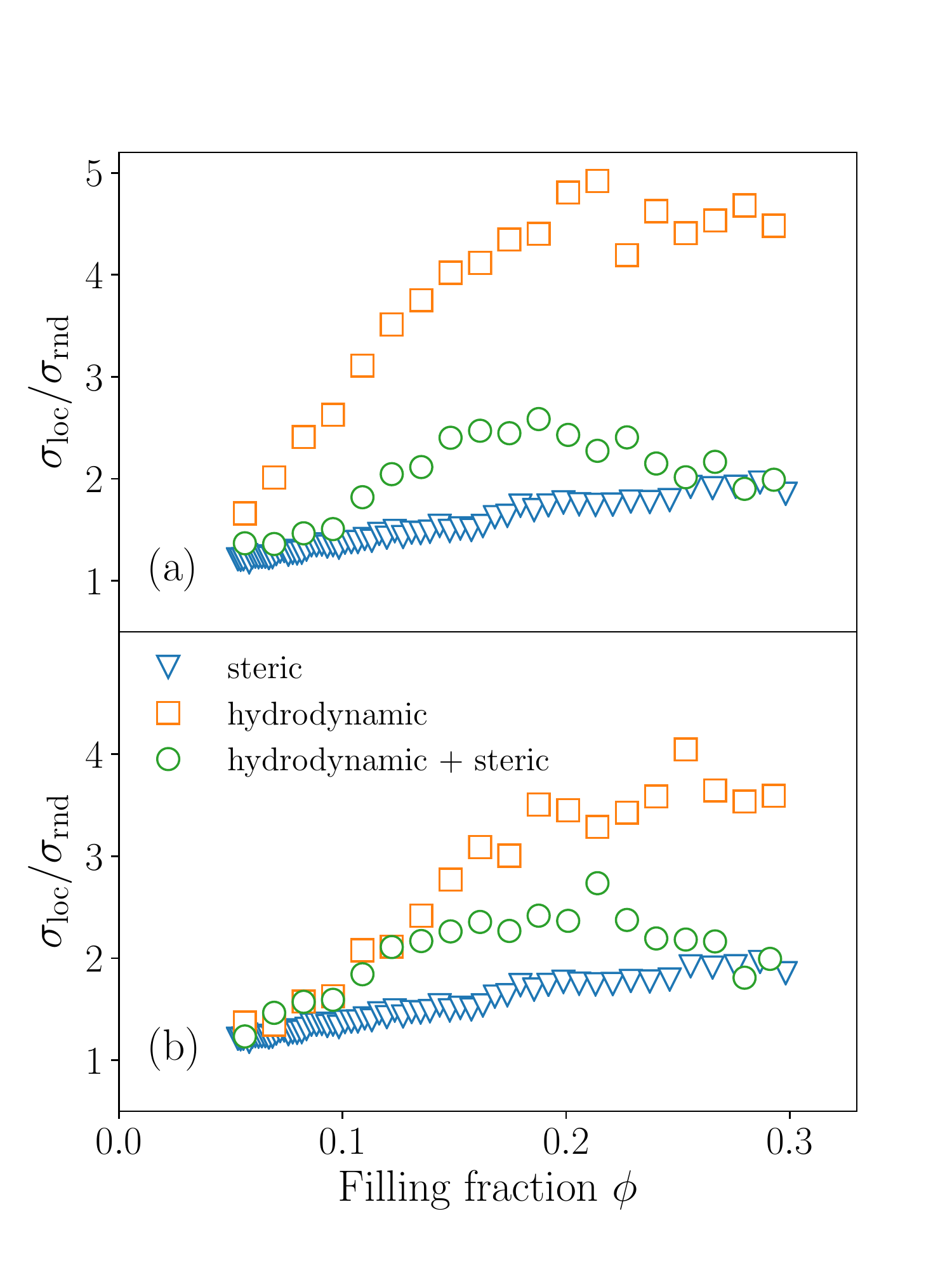}
        \caption{ Standard deviation of local Voronoi volume $\sigma_{\mathrm{loc}}$ compared to 
        standard deviation $\sigma_{\mathrm{rnd}}$ of a homogeneous configuration. Global filling fraction is varied and the P\'eclet number is fixed to $\mathcal{P} = 2.6 \times 10^3$.
                Panel (a) show pusher and (b) puller type swimmers. 
                The circles are simulations with hydrodynamic and steric interactions, 
                the boxes are simulations including only hydrodynamic interactions,
                and the triangles are simulations including only steric interactions.}
                \label{fig:ComparisonBrownianNosteric}
\end{figure}

To test the hypothesis that steric interactions stabilize the hydrodynamic instability,
we carry out two more types of simulations: first, we exclude the steric effects 
by setting the potential [Eq.~\eqref{eq:WCAPotential}] to zero, thus only hydrodynamic effects are included.
Second, we carry out Brownian dynamics simulations, which completely neglect hydrodynamic interactions.
More details about the Brownian dynamics simulations are given in App.~\ref{subsec:BDsims}.

In Fig.~\ref{fig:ComparisonBrownianNosteric} we show our original simulations that fully account for hydrodynamic and steric interactions, 
the simulations without steric interactions, and the active Brownian simulations.
The simulations with hydrodynamics alone give rise to a strong increase of $\sigma_{\mathrm{loc}}/\sigma_{\mathrm{rnd}}$, but no maximum occurs,
while the active Brownian simulations show an increase of $\sigma_{\mathrm{loc}}/\sigma_{\mathrm{rnd}}$, which is much less pronounced.
The original simulations are at an intermediate regime. Thus, we conclude that the maximum which we see is 
mediated by an interplay of the hydrodynamic interactions and the steric interactions, confirming our hypothesis.

\section{Theoretical Analysis} 
\label{Sec:Analytical}
To bolster our numerical results, we develop an analytical theory of microswimmers that explicitly includes hydrodynamic and steric interactions.
As suggested in \cite{BaskaranPNAS2009} swimmers are modeled by an asymmetric dumbbell, similar to our numerical model (see Sec.~\ref{Sec:model}), where the motion of the swimmer is described
by the following effective Langevin equations
\begin{align}
        \frac{\text{d} \bm{r}_{L i}}{\text{d}t} &= \bm{v}\left( \bm{r}_{L i} \right),
        \label{eq:langevinfront}
        \\
        \frac{\text{d} \bm{r}_{S i}}{\text{d}t} &= \bm{v}\left( \bm{r}_{S i} \right),
        \label{eq:langevinback}
\end{align}
where $\bm{r}_{L i}$ is the position of the front sphere of the swimmer $i$ and $\bm{r}_{S i}$ the position of 
the respective back sphere. 
The front and back spheres of each swimmer are connected by an infinitely thin rigid rod.
The motion is coupled to the fluid velocity $\bm{v}$, which is determined 
by the Stokes equation including a stochastic and an active force term
\begin{align}\label{eq:StokesBaskaran}
\eta\nabla^2\bm{v}=\nabla p-\bm{f}^{\mathrm{active}} + \bm{f}^{\mathrm{noise}}.
\end{align}
Here, the active force is given by a force dipole
\begin{align}
        \bm{f}^{\mathrm{active}} = \sum_{i} f \bm{e}_{i}\left[ \delta\left( \bm{r} -\bm{r}_{L i} \right) -  \delta\left( \bm{r} -\bm{r}_{S i} \right) \right],
        \label{eq:forcelangevineq}
\end{align}
which points along the orientation of the swimmer $\bm{e}_{i}$ and has a force strength $f$. 
Furthermore, fluctuations in the swimmers motion are added to the fluid via
\begin{align}
        \bm{f}^{\mathrm{noise}}=  \sum_{i} \bm{\xi}^L_{ i} (t) \delta\left( \bm{r} -\bm{r}_{L i} \right) - \bm{\xi}^S_{ i}(t) \delta\left( \bm{r} -\bm{r}_{S i} \right),
        \label{eq:StokesFluctuations}
\end{align}
where $\bm{\xi}^{L,S}_{i}(t)$ are noise terms with $\langle \bm{\xi}^{L,S}_{i}(t)\bm{\xi}^{L,S}_{ \beta}(t') \rangle = 2 \Gamma_{L,S} \mathbf{I} k_{\mathrm{B}} T \delta_{ i \beta} \delta(t-t')$,
and $\Gamma_{L,S}= 6\pi \eta a_{L,S}$ are the friction coefficients of the front and back sphere.
Following \cite{BaskaranPNAS2009}, we derive the one-body Smoluchowski equation from Eq.~\eqref{eq:langevinfront}-\eqref{eq:StokesFluctuations}.
As we have seen in Sec.~\ref{Sec:Denshetero} steric interactions among swimmers may play a crucial role in their dynamics. 
Thus, we explicitly include the effect of steric interactions by means of the Ansatz \cite{BialkeEPL2013, StenhammerPRL2013}
\begin{align}
        v(c)= v_0 - c \zeta.
        \label{eq:VelocityDependenceOnRho}
\end{align}
Here, $v_0$ is the bare swimming velocity and $c$ is the concentration.
The constant $\zeta$ quantifies how much the swimmers are slowed down by the steric interactions (for more details see \cite{BialkeEPL2013, StenhammerPRL2013}).
The Smoluchowski equation then reads
\begin{align}
        \partial_t p = &- \nabla \cdot \left( v(c) \bm{e} p \right) - \frac{1}{\zeta_{hy}} \nabla \cdot \left( \bm{F}_{hy} p \right) 
        \nonumber \\ &- \frac{1}{ \zeta_{hy} l^2} \left( \bm{e}\times \frac{\partial}{\partial\bm{e}} \right) \cdot \bm{\tau}_{hy}p  
        \nonumber \\ &+ D \Delta p + D_R \left( \bm{e}\times \frac{\partial}{\partial\bm{e}} \right)^2 p,
        \label{eq:ModifiedSmulchowski}
\end{align}
where $p(\bm{r},\bm{e},t)$ is the one-particle probability distribution function of 
finding a swimmer at position $\bm{r}$, with orientation $\bm{e}$ at time $t$. 
The first term on the right hand side of Eq.~\eqref{eq:ModifiedSmulchowski} takes into account the active 
motion with a density dependent velocity, due to steric interactions;
the second term accounts for the hydrodynamic forces and the third term for the 
hydrodynamic torques; the last two terms are responsible for diffusivity, 
with translational diffusion constant $D$ and rotational diffusion constant $D_R$.
To make progress with this equation, we follow the standard path and compute moment equations for 
the concentration $c$, the polarization $\bm{P}$ and the nematic order tensor $\mathbf{Q}$
\begin{align}
        c\left( \bm{r},t \right) &= \int \text{d}\bm{e} \ p(\bm{r},\bm{e},t), \\
        \bm{P}\left( \bm{r},t \right) &= \frac{1}{c\left( \bm{r},t \right)} \int \text{d}\bm{e} \ \bm{e} \ p(\bm{r},\bm{e},t),  \\
        \mathbf{Q} \left( \bm{r},t \right) &= \frac{1}{c\left( \bm{r},t \right)} \int \text{d}\bm{e} \left( \bm{e} \otimes \bm{e} -\frac{1}{3} \mathbf{I} \right) p(\bm{r},\bm{e},t) .
        \label{eq:MomentDefinition}
\end{align}
The full equations for $c$, $\bm{P}$ and $\mathbf{Q}$ are given in App.~\ref{subsec:HydrodynamicTerms}.
We linearize these moment equations [Eq.\eqref{eq:ConcentrationFull}-\eqref{eq:StressTensorFull}] around the isotropic state, described by $c= c_0 + \delta c$,
$\bm{P} = \delta \bm{P}$ and $\mathbf{Q} = \delta \mathbf{Q}$ and turn to Fourier space,
with wave vector $\bm{k}$, where the fields are denoted by $\delta \tilde c $, $ \delta \tilde{ \bm{P}}$, and $\delta \tilde{ \mathbf{Q}}$. 
The moment equations can then be written in terms of the following fields: the concentration fluctuations $\delta \tilde c$, 
the longitudinal polarization fluctuations $\delta P_{\parallel} = \hat{\bm{k}} \cdot \delta \tilde{ \bm{P}} $, the transverse 
polarization fluctuations $\delta \bm{P}_{\perp} = \delta \tilde{ \bm{P}} \cdot \left( \bm{\delta} - \hat{\bm{k}}\hat{\bm{k}} \right)$,
the ``splay'' $\delta Q_{\parallel \parallel} = \hat{\bm{k}} \cdot \delta \tilde{ \mathbf{Q}} \cdot \hat{\bm{k}} $ and the 
``bend'' $\delta \mathbf{Q}_{\parallel \perp} = \hat{\bm{k}} \cdot \delta \tilde{ \mathbf{Q}} \cdot \left( \bm{\delta} - \hat{\bm{k}}\hat{\bm{k}} \right)$ components of the nematic tensor.
To first order in the fluctuations the equations governing the temporal evolution read
\begin{align}
        \partial_t \delta \tilde c = &\gamma_1 f c_0^2 \delta Q_{\parallel \parallel} + i k (v_0 c_0 - \zeta c_0^2) \delta P_{\parallel} - k^2 D \delta \tilde c,
        \label{eq:LinearConcentration}
        \\
        \partial_t \delta P_{\parallel} = &-D_R \delta P_{\parallel} - k^2 D \delta P_{\parallel}  
                    \nonumber \\  &- \frac{ik}{3} \left( v_0 - 2 \zeta c_0 \right)  \delta \tilde c  - i k \left( v_0 c_0 - \zeta c_0^2 \right)\delta Q_{\parallel \parallel},
        \label{eq:LinearLongitudinalPolarisation}
        \\
        \partial_t \delta \bm{P}_{\perp} = &-D_R \delta \bm{P}_{\perp}  - k^2 D \delta \bm{P}_{\perp} 
                    \nonumber \\  &- \frac{i \bm{k} }{3} \left( v_0 - 2 \zeta c_0 \right)  \delta \tilde c  - i k \left( v_0 c_0 - \zeta c_0^2 \right)\delta Q_{\parallel \perp},
        \label{eq:LinearTransversalPolarisation}
        \\
        \partial_t \delta Q_{\parallel \parallel} = &- 4 D_R \delta Q_{\parallel \parallel} - D c_0 k^2 \delta Q_{\parallel \parallel}  
                    \nonumber \\  &- k^2 D \frac{\delta \tilde c}{c_0} - i k \frac{4}{15} \left( v_0 c_0 - \zeta c_0^2 \right) \delta P_{\parallel},
        \label{eq:LinearSplay}
        \\
        \partial_t \delta \mathbf{Q}_{\parallel \perp} = &- 4 D_R \delta \mathbf{Q}_{\parallel \perp} + \gamma_2 f c_0 \delta \mathbf{Q}_{\parallel \perp} 
                    \nonumber \\  &- D c_0 k^2 \delta \mathbf{Q}_{\parallel \perp} - i k \frac{1}{10} \left( v_0 c_0 - \zeta c_0^2  \right) \delta \bm{P}_{\perp} .
        \label{eq:LinearBend}
\end{align}
Here, $k= |\bm{k}|$ is the absolute value of the wavevector.
The constants $\gamma_1 f$ and  $\gamma_2 f$ (see App.~\ref{subsec:HydrodynamicTerms})
stem from the hydrodynamic interactions between the swimmers, which are dominated by the force dipole strength $f$
and can cause instabilities in the system.
To analyze the stability of the system we will first consider pullers ($f<0$) and in the second step pushers ($f>0$). 

For pullers ($f<0$) the bend fluctuations $\delta \mathbf{Q}_{\parallel \perp}$ are stable. 
To analyze the fluctuations in the concentration we use a large length-scale and long time-scale approximation for the 
longitudinal polarization and the splay component of the nematic order tensor
\begin{align}
         \delta P_{\parallel} \approx &- \frac{ik}{3 D_R} \left( v_0 - 2 \zeta c_0 \right)  \delta \tilde c ,
        \label{eq:LongitudinalPolarisationApprox}
        \\
        \delta Q_{\parallel \parallel} \approx &- k^2 \frac{D}{4 D_R} \frac{\delta \tilde c}{c_0} - i k \frac{1}{15 D_R} \left( v_0 c_0 - \zeta c_0^2 \right) \delta P_{\parallel}.
        \label{eq:BendApprox}
\end{align}
Inserting Eq.~\eqref{eq:LongitudinalPolarisationApprox}-\eqref{eq:BendApprox} 
into the Eq.~\eqref{eq:LinearConcentration} and keeping terms up to $\mathcal{O}(c_0^2)$ yields
\begin{align}
        \partial_t \delta \tilde c = - k^2 &\left[ D + \left( \frac{\gamma_1 D}{4 D_R} f - \frac{v_0^2}{3D_R} \right) c_0 
          + \frac{v_0 \zeta}{D_R}  c_0^2  \right] \delta \tilde c.
        \label{eq:ConcentrationApprox}
\end{align}
Since for pullers $f<0$ the term $\frac{\gamma_1 D}{4 D_R} f c_0 $ introduces an instability at low concentrations $c_0$, which are counteracted by the  
term $ \frac{v_0 \zeta}{D_R}  c_0^2  $, which stabilizes the system at higher concentrations. Since the first 
term stems from the hydrodynamic interactions and the second term from the steric interactions, we can draw the same conclusion as from 
the simulations: the hydrodynamic interactions cause heterogeneities in the system which are suppressed by the steric effects at larger $c_0$. 
Moreover, we find a maximum of instability as a function of $c_0$ and, hence, 
a maximum heterogeneity.

For pushers ($f>0$) the splay fluctuations are stable, whereas the bend fluctuations become unstable. 
In the same large length and time scale limit the transverse polarization becomes
\begin{align}
        \delta \bm{P}_{\perp} = - i k \frac{1}{D_R} \left( v_0 c_0 - \zeta c_0^2 \right) \delta \mathbf{Q}_{\parallel \perp}.
        \label{eq:TransversalPolarisationApprox}
\end{align}
Inserting this into the Eq.~\eqref{eq:LinearSplay} gives
\begin{align}
        \partial_t \delta \mathbf{Q}_{\parallel \perp} =  - &\Biggl[ 4 D_R  + ( D  k^2 - \gamma_2 f  ) c_0
         \nonumber \\ & +  k^2 \frac{1}{10 D_R} \left( v_0^2 c_0^2 - 2 v_0 \zeta c_0^3 +\zeta^2 c_0^4  \right) \Biggl] \delta \mathbf{Q}_{\parallel \perp}.
        \label{eq:SplayApprox}
\end{align}
Here, the term $- \gamma_2 f c_0$ destabilizes the system at low concentrations $c_0$ through bend fluctuations, which are counteracted by the term 
$ k^2 \frac{1}{10 D_R} v_0^2 c_0^2$, that stabilizes the system for higher concentrations. Again, the first term, which destabilizes the system,
stems from the hydrodynamic interactions, whereas the second, stabilizing term comes from the steric interactions. Considering only terms
up to  $\mathcal{O}(c_0^2)$, we also find a maximum of instability, which translates to a maximum heterogeneity in the system.

\section{Conclusions}
\label{Sec:conclusion}

We have presented a new model for biological microswimmers that is
based on Stokeslets and the stroke averaged motion of their flagella.
The Stokeslets were distributed to model the flow fields of \textit{Chlamydomonas} or \textit{E. coli} cells. 
Furthermore, our model takes into account the anisotropic shape of a microswimmer.
Typical for this is the shape of a \textit{Chlamydomonas} cell, which is well modeled by a dumbbell \cite{OstapenkoarXiv2016}. 
Our model allows for great flexibility in the geometrical modeling of microswimmer shapes.
Self-propulsion is generated through a symmetry breaking due to the asymmetric shape and force free motion. 
The fluid is explicitly included with MPCD.

We show that the flow fields produced in our simulations can be predicted using simple formulae from the literature. 
These formulae also correspond to the experimentally measured flow fields \cite{DrescherPNAS2011,DrescherPRL2010}. 
Additionally, we test the effective velocity of the microswimmer model, and find that it depends linearly on the applied force. 

We study the phase diagram in terms of filling fraction and P\'eclet number.
We find that both pullers and pushers exhibit density heterogeneities. 
The density heterogeneities show a maximum at intermediate filling fractions and high P\'eclet number.
To determine the mechanism underpinning this phenomenon, we perform additional simulations, in which either the steric interactions or the hydrodynamic interactions were switched off.
Simulations with active Brownian particles showed a small linear increase in the density heterogeneities,
while simulations without the steric interactions show strong density heterogeneities. 
This is an instability caused only by the hydrodynamic interactions, which is known from the literature \cite{BaskaranPNAS2009, ishikawajfm2008, EzhilanPoF2013, SaintillanPRL2008, UnderhillPRL2008}.
However, no maximum arises in the simulations with only hydrodynamic or only steric interactions,
which shows that the maximum in the density heterogeneities is mediated by an interplay of the 
hydrodynamic and the steric interactions. 
The hydrodynamic interactions destabilize the system, whereas the steric interactions stabilize the system as the filling fraction grows
and thus a maximum in density heterogeneity arises.

% But we found a new effect, namely a maximum density heterogeneity at high P\'eclet number and intermediate filling fractions.
%Using both simulations and analytical theory we showed that this maximum arises because 
%steric interactions stabilize the system and thus counter act to the instability introduced by the hydrodynamic interactions.
Additionally, we include steric as well as hydrodynamic interactions into an analytical theory, based on 
a Smoluchowski equation. We computed the hydrodynamic moments of this equation and performed a linear stability analysis of the moment equations around the homogeneous sate.
For both puller and pusher-type swimmers, we found that at low concentration the system is destabilized by hydrodynamic interactions.
At higher concentrations the instabilities are counteracted by the steric interaction. 
This gives rise to a maximum in the instability of the homogeneous state, 
and thus a maximum heterogeneity in the concentration of swimmers. 

The pictures from both simulations and analytical theory fit together: both
show that the homogeneous state is not stable and there is a maximum of instability.
Also, both analyses show that the instability arises from hydrodynamic interactions and
is suppressed by the steric interactions.

The maximum might have important biological implications:
it means that there is an optimal 
filling fraction and P\'eclet number for the formation of heterogeneous structures.
Bacteria or microalgae exhibiting these optimal parameters are more likely to
form colonies or biofilms.

\section*{Conflicts of interest}
There are no conflicts of interest to declare. 

\section*{Acknowledgements}
We gratefully acknowledge insightful conversations with Johannes Blaschke, Jens Elgeti, Stephan Herminghaus and Kuang-Wu Lee.  F.J.S. gratefully
acknowledges support from the Deutsche Forschungsgemeinschaft (SFB 937, project A20).

\appendix
\section{Center of mass and moment of inertia}
\label{subsec:CoM}
In the body frame, the swimmer is aligned with the $z$ direction and the coordinates of the $B$ and $F$ spheres are $z_1$ and $z_2$, respectively.
Given a homogeneous mass distribution the center of mass of the swimmer is given by
\begin{align}
z_{\text{CoM}} = \left( V_1 z_1 + V_2 z_2  
                - V_{Sc_1} \left( z_1 + \frac{3}{4} \frac{\left( 2 R_1 - h_1 \right)^2}{3 R_1 - h_1} \right) \right.             \nonumber  \\
        \left. - V_{Sc_2} \left( z_2 + \frac{3}{4} \frac{\left( 2 R_2 - h_2 \right)^2}{3 R_2 - h_2} \right) \right)    \nonumber  \\
                \left( V_1 + V_2 - V_{Sc_1} - V_{Sc_2} \right)^{-1} .
    \label{eq:CenterOfMass}
\end{align}
Where $V_i$ are the volumes of the spheres and $V_{Sc_i}$ are the volumes of their spherical caps, which are cut by the other sphere \cite{Bronstein1979}
\begin{align}
        V_{Sc_i}= \frac{1}{3} \pi h_i^2 \left( 3 R_i -h_i \right).
        \label{eq:VolumeSphericalCap}
\end{align}
With the height $h_i$ of the spherical caps
\begin{align}
        h_1 &= \frac{\left( R_2 - R_1 +l \right) \left( R_2 + R_1 -l \right)}{2l}\,,
        \nonumber
        \\
        h_2 &= \frac{\left( R_1 - R_2 +l \right) \left( R_1 + R_2 -l \right)}{2l}.
        \label{eq:HeightSphericalCaps}
\end{align}
The moment of inertia for a spherical cap in the $x$- as well as $y$-direction is
\begin{align}
        I_{Sc_i , (x,y)} &= \rho \int_V \left(x^2 + z^2\right) \text{d}V
        \nonumber \\
        = &\rho \int_0^{2\pi}\text{d}\varphi \int_0^{\text{acos} \left( \frac{R_i-h_i}{R_i} \right) }\text{d}\theta \text{sin} \theta 
        \nonumber \\ 
          & \int_{ \frac{R_i -h_i}{\text{cos}\theta}}^{R_i} \text{d}r 
        r^2 \left[ \left( \text{cos}\varphi \text{sin}\theta r \right)^2 + \left( \text{cos}\theta  r\right)^2 \right]
        \nonumber \\
        = &\rho \pi \frac{1}{60} h_i^2 ( - 9 h_i^3 + 45 h_i^2 R_i - 80 h_i R_i^2 + 60 R_i^3).
        \label{eq:MomentOfInertiaCapxy}
\end{align}
In the $z$-direction it is
\begin{align}
        I_{Sc_i , z} &= \rho \int_V \left(x^2 + y^2\right) \text{d}V
        \nonumber \\
        = &\rho \int_0^{2\pi}\text{d}\varphi \int_0^{\text{acos} \left( \frac{R_i-h_i}{R_i} \right) }\text{d}\theta \text{sin} \theta 
        \nonumber \\ 
        &  \int_{ \frac{R_i -h_i}{\text{cos}\theta}}^{R_i} \text{d}r 
        r^2 \left[ \left( \text{cos}\varphi \text{sin}\theta r \right)^2 + \left( \text{sin}\varphi \text{sin}\theta  r\right)^2 \right]
        \nonumber \\
        = &\rho \pi \frac{1}{30} h_i^3 ( 3 h_i^2 - 15 h_i R_i + 20 R_i^2)\,.
        \label{eq:MomentOfInertiaCapz}
\end{align}
By using the moment of inertia of a sphere $I_{Sp_i}= \frac{8}{15} \rho \pi R_i^5 $
and with the use of the parallel axis theorem,
the moments of inertia of the swimmer are
\begin{align}
        &I_{(x,y)} = I_{Sp_1, (x,y)} - I_{Sc_1, (x,y)} + \rho \left( V_1 - V_{Sc_1} \right) x_1^2
        \label{eq:MomentOfInertiaxy}
        \\
        &I_z = I_{Sp_1,z} + I_{Sp_2,z} - I_{Sc_1,z} - I_{Sc_2,z}.
        \label{eq:MomentOfInertiaz}
\end{align}

%\subsection{Compressability test}
%\begin{figure}[!htbp]
%        \centering
%        \scalebox{0.5}{\includegraphics{comp_test.pdf}}
%        \caption{Compressability test.}
%        \label{fig:ComprTest}
%\end{figure}

\section{Quaternion formulae}
\label{subsec:QuatFormulae}

Quaternions are represented as $\bm{q}=q_0+q_1\bm{i}+q_2\bm{j}+q_3\bm{k}$, with $q_0,\dots,q_3\in \mathbb{R}$, and $\bm{i}^2=\bm{j}^2=\bm{k}^2=\bm{ijk}=-1$.
The quaternion matrix $\mathbf{W}$ is (see also \cite{Allen1987})
\begin{align}
        \mathbf{W}\left( \bm{q} \right) = 
       \begin{pmatrix}
  q_0 & -q_1 & -q_2 & -q_3 \\
  q_1 & q_0 & -q_3 & q_2 \\
  q_2 & q_3 & q_0 & -q_1 \\
  q_3 & -q_2 & q_1 & q_0 \\
 \end{pmatrix} .
        \label{eq:Qmatrix}
\end{align}
The unitary matrix $\mathbf{D}$ that transforms vectors from the lab to the body frame is (see also \cite{Allen1987})
\begin{align}
       \begin{pmatrix}
    q_0^2 + q_1^2 -q_2^2 - q_3^2   &    2\left( q_1 q_2 + q_0 q_3 \right)    &     2\left( q_1 q_2 - q_0 q_2 \right) \\
    2\left( q_2 q_1 -q_0 q_3 \right)       &    q_0^2 - q_1^2 + q_2^2 - q_3^2    &    2\left( q_2 q_3 + q_0 q_1 \right) \\
    2\left( 2 q_3 q_1 + q_0 q_2  \right)    &    2\left( q_3 q_2 - q_0 q_1 \right)    &  q_0^2 - q_1^2 - q_2^2 + q_3^2 \\
 \end{pmatrix} .
        \label{eq:LabBodyTrafo}
\end{align}

\section{Brownian dynamics simulations}
\label{subsec:BDsims}
The Brownian dynamics simulations are carried out with hard spheres, that propel 
forward with a typical speed $v_0$ along their orientation $\bm{e}$ [See also \cite{WysockiEPL2014,FilyPRL2012,RednerPRL2013,BialkePRL2012}]. The equation governing the 
translational motion for the position $\bm{r}$ reads
\begin{align}
        \frac{\text{d} \bm{r}}{ \text{d} t} = v_0 \bm{e} + \bm{F}/\gamma + \bm{\eta},
        \label{eq:BDtranslational}
\end{align}
where $\bm{F}$ is the force between particles and $\bm{\eta}$ is a random white noise 
with zero mean and $\langle \bm{\eta}(t) \bm{\eta}(t') \rangle = 2 D \bm{1} \delta(t-t') $. 
Here, $D= k_{\mathrm{B}} T/ \gamma$ is the translational diffusion constant, which is related to the
friction coefficient $\gamma$. The potential between the particles is a Weeks-Chandler-Anderson potential \cite{WeeksJCP1971}
\begin{equation} \label{eq:WCAPotentialsimple}
  \Phi (r_{ij}) = 
            4 \tilde{\epsilon} \left[ \left(\frac{\sigma}{r_{ij}} \right)^{12}-  
                              \left(\frac{\sigma}{r_{ij}} \right)^6 \right] 
            + \tilde{\epsilon}
\end{equation}
if $r_{ij} < 2^{1/6}\sigma$, and $\Phi (r_{ij}) = 0$ otherwise.
Here $r_{ij}\equiv |\bm{r}_{i}-\bm{r}_{j}|$ is the distance between swimmer $i$ and swimmer $j$ and $\tilde{\epsilon} = 1000 k_{\mathrm{B}} T$ is the energy scale.
Furthermore, we include orientational diffusion by using
\begin{align}
        \frac{\text{d} \bm{e}}{ \text{d} t} = \bm{\zeta} \times \bm{e} 
        \label{eq:BDorientational}
\end{align}
where $\bm{\zeta}$ is a Gaussian white noise with $\langle \bm{\zeta}(t) \bm{\zeta}(t') \rangle = 2 D_r \bm{1} \delta(t-t') $.
Here, the rotational diffusion coefficient is related to the translational diffusion coefficient by $D_r= 3 D/\sigma^2 $. 
The P\'eclet number is defined by $\mathcal{P}= v_0 \sigma/D$, equivalently to the definition in Eq.~\eqref{eq:PecletNumber}.

\section{Hydrodynamic terms}
\label{subsec:HydrodynamicTerms}

The moment equations are
\begin{align}
        \partial_t c = &- \nabla \cdot \left( v(c) c \bm{P} \right) + D \Delta c - \nabla \cdot \left( \bm{H}(\bm{r},t) c \right), 
        \label{eq:ConcentrationFull}
        \\
        \partial_t cP_i = &- \partial_j \left( v(c) c Q_{ij} \right) - \frac{1}{3} \partial_i \left( v(c) c \right)  + D \Delta c P_i 
        \nonumber \\ &- D_R c P_i - \nabla \cdot \left( \bm{H}(\bm{r},t) c P_i \right) 
        - H_{2, ij}(\bm{r},t) c P_i 
        \nonumber \\ &+ H_{3,j}(\bm{r},t) c \left( Q_{ij} + \frac{2}{3} \delta_{ij} \right) ,
        \label{eq:PolarisationFull}
        \\
        \partial_t c Q_{ij} = &- \frac{2}{5}\left( \partial_i v(c) c P_j \right)^{\mathrm{ST}} + D \Delta c Q_{ij} - 4 D_R c Q_{ij}
        \nonumber \\ &- \nabla \cdot \left( \bm{H}(\bm{r},t) c Q_{ij}  \right) - \left( H_{4, ij} \right)^{\mathrm{ST}} c 
        \nonumber \\ &- \left( \bm{H}_5 \cdot c \bm{P} \right)_{ij}^{\mathrm{ST}},
        \label{eq:StressTensorFull}
\end{align}
where the $H_{\alpha}$ terms stem from the hydrodynamic interactions and are given in the following.
The superscript $\mathrm{ST}$ stands for the symmetric traceless contraction $[Y_{ij}]^{\mathrm{ST}} = (1/2) (Y_{ij} + Y_{ji}) - (1/3) \delta_{ij} Y_{kk}$.
The terms governing the hydrodynamic interactions in Eq.~\eqref{eq:ConcentrationFull}-\eqref{eq:StressTensorFull} are (see also \cite{BaskaranPNAS2009})
\begin{align}
        &\bm{H}(\bm{r},t) = \frac{\alpha_1}{\zeta_{\mathrm{hy}}} \bm{K}^{F_1} (\bm{r},t)  -  \frac{\beta_1}{\zeta_{\mathrm{hy}}} \bm{K}^{F_2} (\bm{r},t) ,
        \label{eq:HydroFak1} \\
        &H_{2,ij}= \frac{\alpha_2}{5 \bar\zeta l^2} \left( 4 K^{\tau_1}_{ij} (\bm{r},t) -  K^{\tau_1}_{ji} (\bm{r},t) - \delta_{ij}  K^{\tau_1}_{mm} (\bm{r},t)   \right),
        \label{eq:HydroFak2} \\
        &H_{3,j}= \frac{\beta_3}{\bar\zeta l^2} K^{\tau_2}_{j} (\bm{r},t),
        \label{eq:HydroFak3} \\
        &H_{4,ij} = \frac{2 \alpha_2}{5 \zeta_{\mathrm{hy}} l^2} K^{\tau_1}_{ji} (\bm{r},t),
        \label{eq:HydroFak4} \\
        &\bm{H}_5= \frac{\beta_3}{\zeta_{\mathrm{hy}} l^2} \bm{K}^{\tau_2},
        \label{eq:HydroFak5}
\end{align}
where the terms stemming from hydrodynamic forces are
\begin{align}
        &K_i^{F_1}(\bm{r}_1,t) = \int \text{d} \bm{r}_2 \frac{\hat r_{12 i}}{r_{12 i}^2} S_{jk}( \bm{\hat r}_{12}) c(\bm{r}_2,t) Q_{jk}(\bm{r}_2,t),
        \label{eq:HydroKF1}\\
        &K_i^{F_2}(\bm{r}_1,t) = \frac{2}{5} \int \text{d} \bm{r}_2 \frac{1}{r_{12 i}^3} S_{ij}(\bm{\hat r}_{12}) c(\bm{r}_2,t) P_{j}(\bm{r}_2,t),
        \label{eq:HydroKF2}
\end{align}
and the terms stemming from the torques are
\begin{align}
        &K^{\tau_1}_{nk} (\bm{r}_1,t) = \int \text{d}\bm{r}_2 \frac{\hat r_{12n}}{r_{12}^3} S_{ikl}(\bm{\hat r_{12}})  c(\bm{r}_2,t) Q_{jl}(\bm{r}_2,t),
        \label{eq:HydroKT1}\\
        &K_n^{\tau_2}(\bm{r}_1,t) =  \int \text{d} \bm{r}_2 \frac{1}{r_{12 i}^5} S_{nk}(\bm{\hat r}_{12}) c(\bm{r}_2,t) P_{k}(\bm{r}_2,t).
        \label{eq:HydroKF2}
\end{align}
Furthermore
\begin{align}
        &S_{kl}(\bm{\hat r}_{12})= \left[ \hat r_{12k}\hat r_{12k} - \frac{1}{3} \delta_{kl} \right],
        \label{eq:HydroS1}\\
        &S_{ijk}(\bm{\hat r}) = 5 \hat r_i\hat r_j\hat r_k - \left(  \delta_{ik}\hat r_i + \delta_{ik}\hat r_j +\delta_{ij} \hat r_k \right),
        \label{eq:HydroS2}
\end{align}
and the constants are
\begin{align}
        \alpha_1&= \frac{9}{4} f \bar a l, \\
        \alpha_2 &= \frac{9}{4} f l^3 \bar a, \\
        \beta_1 &= - \frac{9}{16} f l^2 \left( a_L - a_S \right), \\
        \beta_2 &= -\frac{9}{16} f l^4 \left( a_L - a_S \right),\\
        \beta_3 &= \frac{9 f l^5}{16 \bar a} \left( a_L - a_S \right)^2,\\
        \zeta_{\mathrm{hy}} &= 6\pi \bar a \eta,
        \label{eq:HydroConstants}
\end{align}
where $a_L$ and $a_S$ are the radii of the front and back sphere, $l$ is their distance and $\bar a= (a_L +a_S)/2$.
The constants accounting for the hydrodynamics in Eq.~\eqref{eq:LinearConcentration}-\eqref{eq:LinearBend} are
\begin{align}
        \gamma_1 = \frac{l}{2 \eta},
        \label{eq:HydrodynGamma1}\\
        \gamma_2 = \frac{3l}{75 \eta}.
        \label{eq:HydrodynGamma2}
\end{align}
%\bibliography{DAS} 
%\bibliographystyle{rsc} %the RSC's .bst file

%merlin.mbs apsrev4-1.bst 2010-07-25 4.21a (PWD, AO, DPC) hacked
%Control: key (0)
%Control: author (8) initials jnrlst
%Control: editor formatted (1) identically to author
%Control: production of article title (-1) disabled
%Control: page (0) single
%Control: year (1) truncated
%Control: production of eprint (0) enabled
%

\end{document}